\documentclass[a4paper,11pt]{article}
\usepackage{pos}
\usepackage[separate-uncertainty = true]{siunitx}
\usepackage{hyperref}
\usepackage[output-decimal-marker={.}]{siunitx}

\newlength{\bibitemsep}
\newlength{\bibparskip}
\let\oldthebibliography\thebibliography
\renewcommand\thebibliography[1]{%
  \oldthebibliography{#1}%
  \setlength{\parskip}{\bibitemsep}%
  \setlength{\itemsep}{\bibparskip}%
}

\title{Deep learning reconstruction of neutrino direction, energy, and flavor with complete uncertainty predictions}
\ShortTitle{Deep learning reconstruction of neutrino direction, energy, and flavor}

\author*[a]{Nils Heyer}
\author[a]{Thorsten Glüsenkamp}
\author[a]{Christian Glaser}

\affiliation[a]{Dept. of Physics and Astronomy, Uppsala University, Box 516, S-75120 Uppsala, Sweden}

\emailAdd{nils.heyer@physics.uu.se}

\abstract{With the IceCube-Gen2 observatory under development and RNO-G under construction, the first detection of ultra-high-energy neutrinos is on the horizon making event reconstruction a priority. Here, we present a full reconstruction of the neutrino direction, shower energy, and interaction type (and thereby flavor) from raw antenna signals. We use a deep neural network with conditional normalizing-flows for the reconstruction. This, for the first time, allows for event-by-event predictions of the posterior distribution of all reconstructed properties, in particular, the asymmetric uncertainties of the neutrino direction. The algorithm was applied to an extensive MC dataset of 'shallow' and 'deep' detector components in South Pole ice. We present the reconstruction performance and compare the two station components. For the first time, we quantify the effect of birefringence on event reconstruction.}

\FullConference{10th International Workshop on Acoustic and Radio EeV Neutrino Detection Activities (ARENA2024)\\
11-14 June 2024\\
The Kavli Institute for Cosmological Physics, Chicago, IL, USA\\}

\begin{document}
\maketitle

\section{Introduction}
During the last decade, there have been tremendous advances in astroparticle physics and high-energy neutrino detection. The IceCube neutrino observatory at the South Pole has successfully measured the cosmic neutrino flux \cite{cosmic} and identified several galactic \cite{galactic} and extra-galactic \cite{TXS_0506, NGC_1068} sources of neutrinos. However, questions remain about even more energetic neutrinos: What is the neutrino flux at ultra-high energies, what are the sources of these neutrinos, and can we measure and reconstruct their properties? One of the most promising approaches to tap into the ultra-high energy neutrino flux is the in-ice radio technique \cite{Barwick:2022vqt}. Exploiting the Askaryan effect from neutrino-induced particle showers and the long attenuation length in polar ice sheets it offers a cost-effective way to instrument enormous volumes of ice. The ARIANNA \cite{ARIANNA1}, ARA \cite{ARA}, and ANITA \cite{anita} experiments have gathered a lot of data and insight into how to build such a detector. The RNO-G \cite{rnog} experiment currently under construction in Greenland and the proposed order of magnitude larger IceCube Gen2-Radio \cite{TDR} experiment planned to be deployed at the South Pole might be the first to detect an ultra-high energy neutrino. 

Measuring an ultra-high energy neutrino is a matter of building a detector with a large enough effective area. To get the most out of these detectors they should not be designed to be pure counting experiments of neutrinos but also capable of characterizing event properties precisely. Many physics applications of ultra-high energy neutrino detectors go beyond measuring the neutrino flux. If the detectors are built such that they can reconstruct the neutrino's energy, direction, and flavor, it would be possible (given enough statistics) to conduct point source searches \cite{point_sources}, constrain source models \cite{flux_pred}, measure flavor compositions \cite{flavor}, and the ultra-high energy neutrino cross section \cite{cross_section}. 

Machine learning, and in particular deep learning, has already helped many physics experiments to get the most out of their measured data, increase the amount of usable data they have available, and even optimize their detector design \cite{optimize}. Traditionally deep learning methods have struggled to not only provide a best-fit-value but also to quantify their uncertainty. New methods such as conditional normalizing-flows \cite{norm_flow, Glusenkamp2024} tackle these shortcomings by providing the full posterior PDF instead of a single best-fit-value. Applying these new methods to in-ice radio detection can help to get the most science out of the new detectors.

\begin{figure}[tbp]
  \centering
  \includegraphics[height=6in]{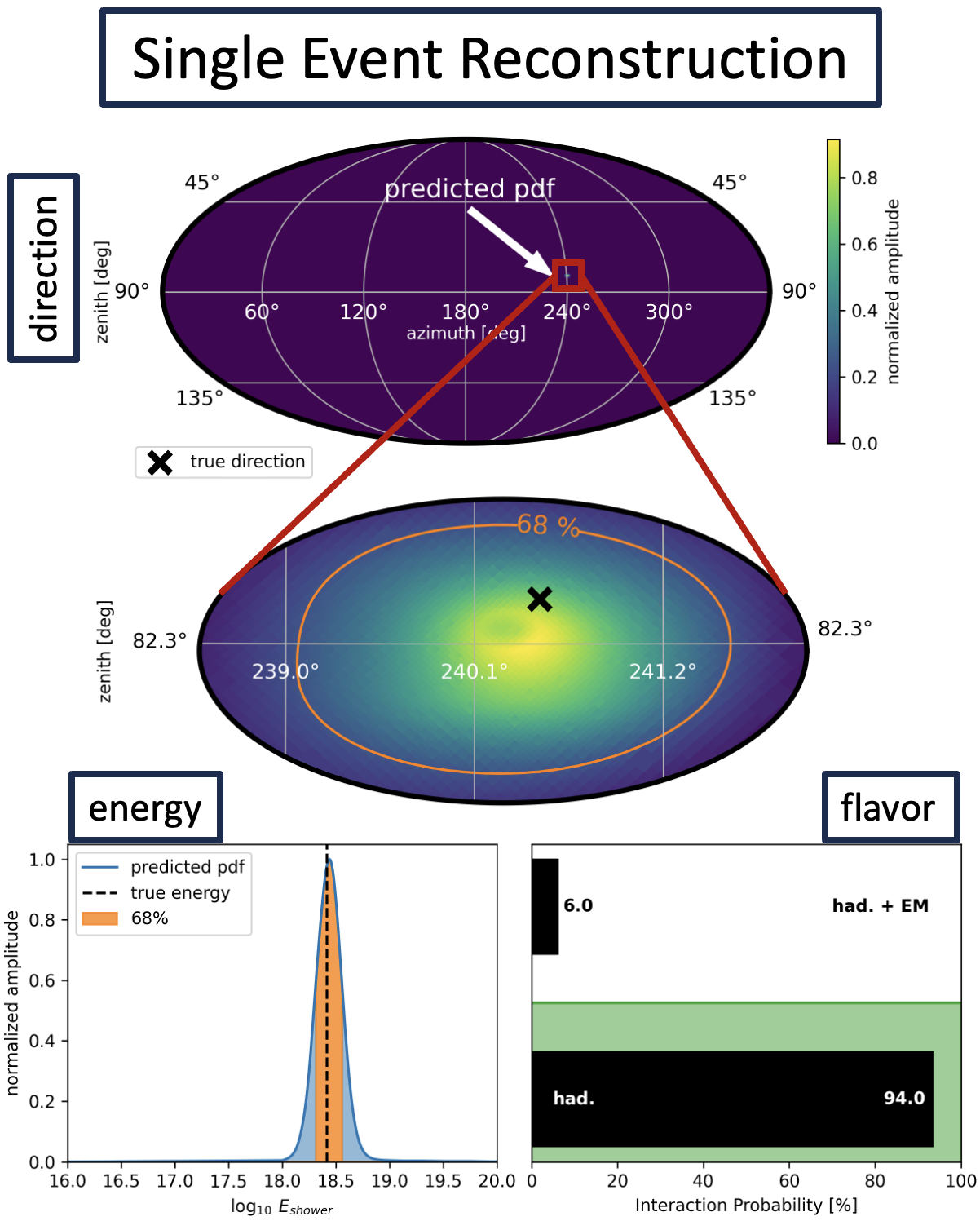}
  \caption{An overview of all properties of the reconstruction for a single event from the test data set. The \textbf{top} two plots show the results from the neutrino direction reconstruction, with the lower of the two being the zoomed-in version of the top one. The green/yellow PDF is the output of the normalizing-flows, the black cross indicates the MC-true direction of the event, and the orange line shows the 68\% contour. The \textbf{lower left} plot shows the results from the shower energy reconstruction. The blue area indicates the PDF predicted by the normalizing-flow, the black dashed line indicates the MC-true shower energy of the event, and the orange area shows the 68\% region. The \textbf{lower right} plot shows the classification results. The black bars are the percentages predicted by the network for the two possible event topologies, and the green region indicates the MC-true event type.}
  \label{fig:overview}
\end{figure}

This contribution aims to show what is possible in terms of reconstruction for current and future in-ice radio neutrino detectors. We show the resolution that can be reached with the current design of an IceCube Gen2-Radio station \cite{TDR} and our newly developed reconstruction algorithm. Using normalizing-flows, we can predict a full posterior distribution of every event for the shower energy and neutrino direction and classify the interaction type which gives a handle on the neutrino flavor. An example of all the reconstructed quantities can be seen in figure \ref{fig:overview}. The developed algorithm allows us to fully quantify the statistical uncertainty of a given neutrino event. For the first time we studied the impact of birefringence effects \cite{heyer} on the reconstruction. We retrained a model for a 'shallow' station on data that includes birefringence effects, showing that they can be corrected for in a reconstruction study. This work is a continuation of a previous reconstruction study \cite{myICRC_deep}.

\section{Monte Carlo Data}
To be able to reconstruct a measured neutrino event, we currently rely on Monte Carlo data to train the algorithm, as no neutrino event has been measured with the radio method yet. The NuRadioMC and NuRadioReco frameworks \cite{NuRadioMC, NuRadioReco} provide a description of the detection process and return signal traces as if they were measured by the antennas after a neutrino interaction. They force a neutrino interaction in the simulated volume, describe the particle shower and the radio emission from the shower up to the radio propagation through the ice, and the response of the antennas to the radio signal. For this study, we simulated two detector designs from the planned IceCube Gen2-Radio detector at the South Pole. The 'shallow' detector component combines four surface LPDA antennas with a single v-pol antenna, sitting at \SI{-10}{\meter} below the surface. The 'deep' detector component combines twelve v-pol and four h-pol antennas in three boreholes from \SI{-45}{\meter} down to \SI{-150}{\meter}. 2.1 million events were simulated for each detector type over a uniform shower energy spectrum from \SI{ e16}{\eV} up to $\SI[parse-numbers=false]{10^{20.2}}{\eV}$, to avoid bias in the most relevant energy region ($\sim$\SI{ e17}{\eV} - \SI{ e19}{\eV}). Two event topologies were simulated based on the shower type of the neutrino-induced interaction. As we consider the signals arriving in a single station, all events inducing a single hadronic shower look the same to the reconstruction. This includes neutral-current interactions (NC) of all neutrino flavors as well as charged current interactions of muon-, and tau-neutrinos because the created muon or tau in a charged current interaction leaves the detection volume of a single station, thus leaving only a hadronic shower behind. The only different topology is the electron neutrino charged current interactions (CC), as here an electromagnetic shower initiated by the electron overlaps with the hadronic shower from the initial neutrino interaction. 'Shallow' detector components were triggered by requiring a time-coincident high/low threshold crossing in two out of four LPDAs and 'deep' detector components were triggered by a phased array consisting of four v-pol antennas. To have unbiased results, the data set was split into three data sets (training, validation, and testing). For the results shown in section \ref{sec:results}, only the test data set, which the algorithm had never seen before, was used. In order to test how birefringence influences the reconstruction all triggered events were re-simulated with ice that includes birefringence effects. 

\section{Deep Learning Algorithm}
For this study, we wanted to design an algorithm that does not have any information about the event topology (hadronic or hadronic plus electromagnetic) but can give back all relevant information about the event properties with uncertainties. To do this we made use of ResNet blocks \cite{resnet} and conditional normalizing-flows \cite{norm_flow, glüsenkamp}. For the ResNet architecture, we took inspiration from a Kaggle challenge about detecting gravitational waves \cite{grav_waves}. For the normalizing-flows we used the \emph{jammy\textunderscore flows} package \cite{jammy} which allows us to integrate the mapping procedure into a PyTorch network.

For each detector component ('shallow' and 'deep') we trained a separate network. For the 'shallow' components, the input data was a \texttt{[5 x 512]} dimensional array (five antennas and 512 time samples at a sampling rate of 2.4 samples per ns) while for the 'deep' component, the input data was a \texttt{[16 x 2046]} dimensional array (16 antennas with 2046 samples at a sampling rate of 2.4 samples per ns). Via one-dimensional CNN layers (2 for the 'shallow' components and 4 for the 'deep' components) the arrays were compressed to an array of the size \texttt{[\# antennas x 256 x 256]} where the last dimension contains different representation filters. From here the data was treated as an image where the colour channels correspond to the number of antennas. We then used a slightly modified version of the ResNet-34 \cite{resnet} architecture with 512 output nodes. The output nodes were then handed to three separate sections for the shower energy, neutrino direction, and shower type reconstruction. For the shower energy, we used a Gaussianization flow for a one-dimensional Euclidean property. This flow returns the posterior PDF over the shower energy where we use sample mean and sample variance to quantify the reconstruction and its uncertainty. For the neutrino direction, we used a spherical-spline-flow for a two-dimensional property on a sphere. This flow returns the posterior over the neutrino direction where we use the mean as the best-fit-value and the area of the 68-percent contour as the uncertainty of the event to quantify the results. For the shower type, we used a fully connected layer to a single node with a sigmoid activation function. The percentage that is returned can be interpreted as the probability that the given event came from a hadronic shower overlapping with an electromagnetic shower, i.e., from an electron neutrino charged-current interaction. 

\section{Resolution} \label{sec:results}
\begin{figure}[tbp]
  \centering
  \includegraphics[height=2.2in]{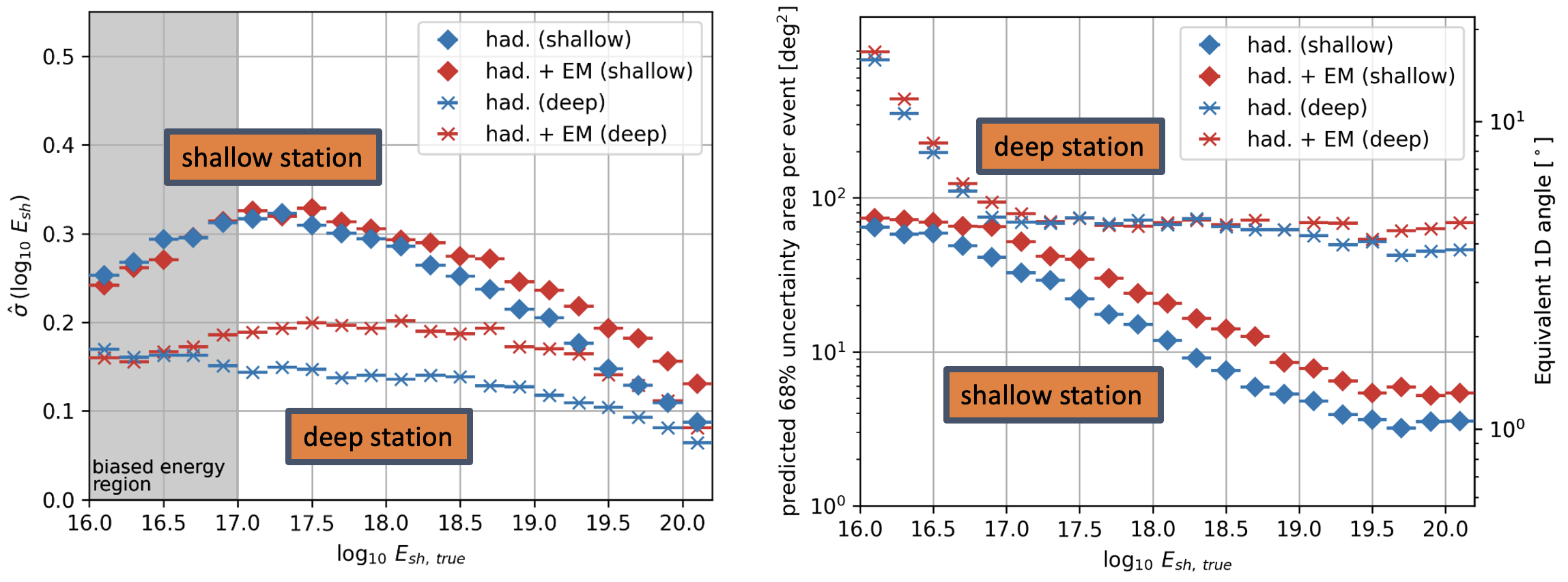}
\caption{Results of the shower energy (\textbf{left}) and neutrino direction (\textbf{right}) reconstruction. Every bin in the plots covers 0.2 in the logarithm of the shower energy, where every bin contains the same number of events ($\sim$5000). The diamonds show the results for the 'shallow' station reconstruction while the crosses show the results from the 'deep' stations. The blue markers are for only hadronic shower events while the red markers are for hadronic and electromagnetic showers. For the shower energy, we plot the median size of the square root of the variance from the predicted distribution on the y-axis. For the neutrino direction, we plot the median size of the orange contours from the top plot of figure \ref{fig:overview} on the left y-axis and the space angle difference that corresponds to this size if the contour was Gaussian on the right y-axis.}
  \label{fig:compare}
\end{figure}

To make a statistically significant statement about the resolution we analyzed the PDF sizes predicted for 100,000 events per station type. Similar to the training data set, the test data set was distributed uniformly in shower energy. Figure \ref{fig:compare} shows the results from the shower energy and the neutrino direction reconstruction. Plotted is the square root of the median variance from the predicted distributions against the shower energy for the energy reconstruction and the median size of the 68\% contour for the direction reconstruction. The plots make it possible to directly compare the resolution of the different event topologies and station layouts. To understand if the predicted uncertainty contours can be trusted we investigated the coverage of the reconstruction, a measure of how many of the 'true' energies/directions lie in the correct uncertainty contour. For the most part, coverage was within 5\% of the expected value. Only some outliers at energies below \SI{ e17}{\eV} had an under coverage down to 20\% for the neutrino direction reconstruction. 

For the shower energy reconstruction, there is a bias visible for both topologies and station layouts between \SI{ e16}{\eV} and \SI{ e17}{\eV} due to which the smaller uncertainties predicted at these energies cannot be trusted. Except for the low energy bias the resolution drops with shower energy. The simpler hadronic showers have significantly smaller uncertainties (up to \SI{0.5}{\eV} in log E) than the more complex showers where an electromagnetic shower overlaps the hadronic shower. The discrepancy gets larger at higher energies, indicating that the difference comes from pulses affected by the LPM effect. The uncertainties from the 'deep' stations are significantly (up to \SI{0.15}{\eV} in log E) smaller than for 'shallow' stations. This can be explained as the 'deep' stations have more antennas and can thus map a bigger portion of the Cherenkov cone compared to the 'shallow' stations. As these results quantify the uncertainty on the shower energy, it is important to mention that for NC events an additional uncertainty of about $\sim$\SI{0.3}{\eV} in log E has to be introduced due to the inelasticity of the event when trying to calculate the neutrino energy. 

For the neutrino direction, again, hadronic showers show smaller contours than the events with an electromagnetic component. The uncertainties from the 'shallow' stations are significantly smaller than for 'deep' stations. This can be explained as the antennas used in 'deep' stations are limited by the diameter of the borehole which limits the ability to measure the horizontal polarization component of the signal. As the polarization is crucial to reconstructing the neutrino direction, this limits the angular resolution. 

The interaction type classification showed a 73\% true positive rate at 10\% false positive rate for correctly identifying electromagnetic shower components for 'deep' stations and a \SI{ e18}{\eV} shower. For 'shallow' stations the true positive rate was about 64\% at the same false positive rate and shower energy. We speculate that the cause of the difference is due to the larger number of antennas with larger spacing. The accuracy of the classification increased significantly with rising shower energy due to the LPM effect.

The resolution of the energy reconstruction was improved substantially compared to similar previous analyses \cite{rnog_energy, Glaser_2023} without using any quality cuts. The resolution of the direction reconstruction was reduced by about a factor of 10 for the 'deep' stations and by about 2 degrees in space angle difference for the 'shallow' stations compared to similar previous analyses \cite{Sjoerd, Glaser_2023}. The results from the flavor classification were comparable to a previous analysis \cite{Oscar, flavor}. We note that these comparisons are not exact measures due to potential differences in the datasets and event selection.

\subsection{The Impact of Birefringence on the Reconstruction}
The ice at the South Pole is known to have birefringent properties \cite{Jordan}. However, the impact this effect will have on the radio detection of neutrinos is still not fully understood. There are studies on how birefringence alters a radio pulse \cite{heyer}, and how it would affect the sensitivity of a radio neutrino detector \cite{icrc_bire}, but so far no study has been done on the impact of birefringence on the achieved resolution of a reconstruction. In previous reconstruction studies, the effect was mentioned as a systematic uncertainty but its effect wasn't quantified. Here we present the first study on the impact of birefringence on the reconstruction. For this purpose, we re-simulated all events from our 'shallow' dataset with the NuRadioMC implementation of birefringence effects \cite{heyer} and repeated the full reconstruction, training a completely new model. Previous studies have shown that birefringence has two major effects on radio pulses: It changes their polarization due to a mixing of polarization Eigenstates, and it introduces a time delay between these states. The magnitude of these effects depends strongly on the azimuth angle by which the pulse propagates through the ice compared to the glacial ice-flow. The intention behind this study was to show that a deep learning approach can pick up on this dependence and still manage to perform the reconstruction. 

\begin{figure}[tbp]
  \centering
  \includegraphics[height=2.2in]{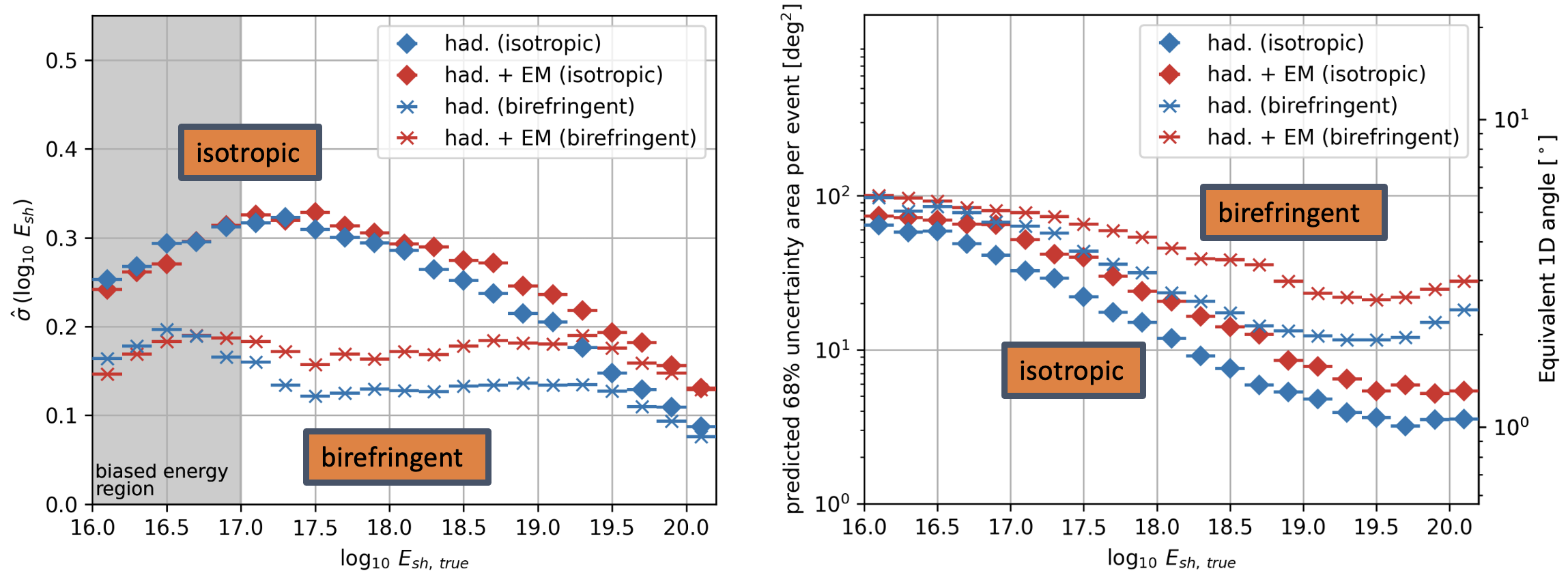}
\caption{Results of the shower energy (\textbf{left}) and neutrino direction (\textbf{right}) reconstruction for birefringent and isotropic data for 'shallow' stations (similar as figure \ref{fig:compare} but with birefringence effects).}
  \label{fig:birefringence}
\end{figure}

The results of the reconstruction can be seen in figure \ref{fig:birefringence}. The direction resolution gets worse when considering birefringence effects which can be explained by a more complex propagation of the radio signal where the polarization of the pulse gets altered, making it harder for the model to reconstruct it. However, the reconstruction is still able to perform on a similar level compared to the results that did not include the birefringence effects. Interestingly the energy, reconstruction improves significantly when considering birefringence effects. We interpret this effect as a result of the introduced time delay between polarization eigenstates. This time delay depends linearly on the distance to the neutrino interaction vertex. By significantly improving the vertex reconstruction, also the energy resolution is improved.

\section{Conclusion}
 For this contribution, we created a large Monte Carlo dataset for in-ice radio stations detecting neutrinos at the South Pole. Using this dataset we trained a neural network to reconstruct the shower energy, neutrino direction, and shower type of the event. For the shower energy and the neutrino direction, we predict the full uncertainty PDF for every event. We evaluated the reconstruction by the size of the predicted uncertainties while checking for sufficient coverage. At \SI{1}{\exa \eV} 'deep' stations have an energy resolution of  0.15 in log E and a direction resolution of 70 square degrees for a simple hadronic shower. At the same energy with the same shower type, the 'shallow' stations have an energy resolution of 0.3 in log E and a direction resolution of 10 square degrees. The Impact of birefringence was investigated for the first time to see if data including the birefringence effect could still be reconstructed. This turned out to be the case with the direction reconstruction suffering slightly while the energy reconstruction improved significantly. We were able to relate the changes in the resolution due to birefringence to the effects of birefringence on radio pulses. The work on the reconstruction is still ongoing with further improvements being investigated. 

\bibliographystyle{ICRC}
\bibliography{references}

\end{document}